\documentclass[11pt, preprint]{aastex631}
\usepackage{rotating}
\usepackage{bm}
\graphicspath{{./}{figures/}}
\newcommand{\ii}{\mathrm{in}}
\newcommand{\oo}{\mathrm{out}}
\newcommand{\rpo}{r_\mathrm{p,\oo}}
\newcommand{\ain}{a_\ii}

\newcommand{\imut}{i_\mathrm{mut}}
\newcommand{\Td}{T_\mathrm{d}}

\usepackage{xcolor}
\definecolor{steelblue}{rgb}{0.275, 0.510, 0.706}
\definecolor{seagreen}{rgb}{0.190, 0.525, 0.361}
\definecolor{alcrimson}{rgb}{0.227, 0.38, 0.54}
\received{2022 September 12}
\revised{2022 December 12}
\accepted{2022 December 14}
\submitjournal{ApJ}
\shorttitle{Lagrange vs. Lyapunov stability of triples}
\shortauthors{Hayashi, Trani, Suto} 
\begin{document}

\title{Lagrange vs. Lyapunov stability of hierarchical triple
  systems:\\ dependence on the mutual inclination between inner and
  outer orbits}
\correspondingauthor{Toshinori Hayashi}
\email{toshinori.hayashi@phys.s.u-tokyo.ac.jp}

\author[0000-0003-0288-6901]{Toshinori Hayashi}
\affiliation{Department of Physics, The University of Tokyo,  
Tokyo 113-0033, Japan}
\author[0000-0001-5371-3432]{Alessandro A. Trani}
\affiliation{Research Center for the Early Universe, School of Science,
The University of Tokyo, Tokyo 113-0033, Japan}
\affiliation{Okinawa Institute of Science and Technology Graduate University,
  Okinawa 904-0495, Japan}
\author[0000-0002-4858-7598]{Yasushi Suto}
\affiliation{Department of Physics, The University of Tokyo,  
Tokyo 113-0033, Japan}
\affiliation{Research Center for the Early Universe, School of Science,
The University of Tokyo, Tokyo 113-0033, Japan}
\affiliation{Laboratory of Physics, Kochi University of Technology,
  Tosa Yamada, Kochi 782-8502, Japan}

\begin{abstract}
  While there have been many studies examining the stability of
  hierarchical triple systems, the meaning of ``stability'' is
  somewhat vague and has been interpreted differently in previous
  literatures.  The present paper focuses on ``Lagrange stability'',
  which roughly refers to the stability against the escape of a body
  from the system, or ``disruption'' of the triple system, in contrast
  to ``Lyapunov-like stability'' that is related to the chaotic
    nature of the system dynamics.  We compute the evolution of
  triple systems using direct $N$-body simulations up to $10^7 P_\oo$,
  which is significantly longer than previous studies (with $P_\oo$
  being the initial orbital period of the outer body). We obtain the
  resulting disruption timescale $\Td$ as a function of the triple
  orbital parameters with particular attention to the dependence on
  the mutual inclination between the inner and outer orbits,
  $\imut$. By doing so, we have clarified explicitly the difference
  between Lagrange and Lyapunov stabilities in astronomical
  triples. Furthermore, we find that the von Zeipel-Kozai-Lidov
  oscillations significantly destabilize inclined triples (roughly
  with $60^\circ < \imut < 150^\circ$) relative to those with
  $\imut=0^\circ$. On the other hand, retrograde triples with
  $\imut>160^\circ$ become strongly stabilized with much longer
    disruption timescales.  We show the sensitivity of the normalized
  disruption timescale $\Td/P_\oo$ to the orbital parameters of triple
  system. The resulting $\Td/P_\oo$ distribution is practically more
  useful in a broad range of astronomical applications than the
  stability criterion based on the Lyapunov divergence.
 \end{abstract}


\section{Introduction \label{sec:intro}} 

Hierarchical triple systems play a very basic and important role in a
broad range of astronomical phenomena. Their dynamical stability has
numerous implications for actual systems
\citep[e.g.,][]{Jha2000,Naoz2013,Antonini2014,
  Ransom2014,Perpinya2019,Rivinius2020,
  Bodensteiner2020,Toonen2020,Tokovinin2020,
  Hayashi2020a,Hayashi2020b}. Furthermore, it remains as one of the
most long-standing questions in celestial mechanics
\citep[e.g.][]{Eggleton1995,Mardling1999,Mardling2001,Georgakarakos2013,Grishin2017,He2018,
  Myllari2018,Wei2021,Lalande2022,Tory2022,Vynatheya2022}.

Among all, the dynamical stability criterion for hierarchical triples
proposed by \citet[][hereafter MA01]{Mardling2001} is well-known and
widely used in various studies of the triple dynamics.  They examine
the stability of triples as follows (R. Mardling, private
communication). First, they derive the analytic expression for the
stability criterion based on the chaotic evolution boundary \citep[see
  also][]{Mardling1995,Mardling1995b}. In order to determine the
numerical factor, they integrate two almost identical systems (the
given system and its ``ghost'') except for their different inner
eccentricities $e_\ii$ of $\Delta e_\ii= 10^{-7}$, and monitor the
difference in the inner semi-major axes $a_\ii$ at the time of outer
apocenter passage.  This variable is expected to be fairly
  insensitive to the initial difference for a stable system, while
grow exponentially for chaotic and unstable systems. This
  consideration motivated MA01 to judge and classify the stable and
unstable systems from the departure of $a_\ii$ evaluated at 100
times the outer orbital period, $P_\oo$ \citep[see also][]
{Mardling2008}.

 The Lyapunov stability is mathematically defined as the stability
  of solutions of dynamical systems near their fixed (equilibrium)
  points, but has been also interpreted as the stability around
  arbitrary solutions (or trajectories) of dynamical systems against
  the tiny changes of system parameters \citep[see, for
    instance][]{Lichtenberg1983,Kandrup1990,Suto1991,Lichtenberg1992}.
  The methodology of MA01 is devised to identify the chaotic nature of
  triple systems in the spirit similar to the Lyapunov stability in the
  latter sense. We note that their methodology is close to the Lyapunov exponent method, which is widely used to characterize the chaoticity of systems. Therefore, we refer their stability condition as
  the Lyapunov stability below in a broader sense.

 In any case, the Lyapunov stability or the local chaoticity
  may not be directly used in judging the global fate of the
  astronomical triple systems in reality. Even if triple systems are
  locally chaotic, their orbital configurations may be bounded, {\it
    i.e.}, Lagrange stable.  Indeed, the latter stability would
  correspond to a more relevant and practical definition in most
  applications for aircrafts and nuclear-plants
  \citep{Gyftopoulos1963}, as well as for astronomical triple systems
  of our interest.

It seems that the stability criterion of triples derived by MA01
  is sometimes misunderstood, and used as a criterion of triple
  disruption in the sense of Lagrange instability. Recently
  \citet[][hereafter Paper I]{HTS2022} computed the disruption
  timescales $\Td$ of triples by directly integrating the three-body
  dynamics up to $10^9 P_\ii$, where $P_\ii$ is the initial orbital
  period of the inner binary.  The distribution of the normalized
  disruption timescale $\Td/P_\ii$ follows an approximate scaling
  similar to the MA01 criterion for coplanar-prograde triples in which
  the mutual inclination between inner and outer orbits $\imut$ is
  $0^\circ$. The orthogonal and coplanar-retrograde triples, however,
  exhibit very different behavior, illustrating the clear difference
  between Lagrange and Lyapunov stabilities. In particular, Paper I
  finds that the von Zeipel-Kozai-Lidov (ZKL) oscillations
  \citep{Zeipel1910,Kozai1962,Lidov1962} play a major role in
  disrupting significantly inclined triples.  Because the
  characteristic ZKL timescale $\tau_{\rm ZKL}$ is comparable or even
  longer than $100\, P_\oo$ in general, their Lagrange stability
  cannot be determined from the short-term local behavior that
  characterizes Lyapunov stability.

 The relation between the chaoticity and Lagrange stability is
    subtle. Figure 6 of Paper I illustrated such examples; the lower
    panels in the figure show that a Lagrange-unstable triple with
    $\Td \approx 3000 P_\ii$ is robust against the change of the input
    initial conditions, unlike the other two cases (the upper and
    middle panels). Therefore, Paper I concluded that a
    Lagrange-unstable triple is not necessarily sensitive to the
    initial conditions. More strictly, a Lagrange-unstable triple may
    become chaotic eventually (just at the last moment of the
    disruption), but it is not clear how long it takes for them to
    exhibit noticeable chaoticity. Indeed, this is what we intend to
    examine quantitatively in Paper I and the present paper; we record
    the disruption timescales during the simulations, not merely the
    stable/unstable outcome alone.

 In addition, we would like to emphasize that a chaotic triple
    does not always necessarily become Lagrange unstable.  Thus, the
    Lagrange instability cannot be judged simply from the first few
    hundred (or fewer) orbits of a trajectory to its nearby ghost as
    MA01 assumed. This is exactly why we have performed a
    significantly longer time-integration (up to $10^9 P_\ii$ in Paper
    I and $10^7 P_\oo$ in the present paper) to check the Lagrange
    instability of various triples. Only with such direct confirmation
    of the disruption timescales, one can understand the relation
    between the Lagrange instability and the chaotic behavior of the
    triple systems. 

  This point is also discussed in a different approach
    recently by \citet{Gajdos-Vanko2023}, who found that several
    observed exoplanetary systems have relatively short Lyapunov
    times (calculated from the Lyapunov exponent), even down to $O(10)$ yrs. This fact implies that the Lyapunov
    time is not necessarily related to the Lagrange stability of the
    systems, in good agreement of our finding mentioned in the above.

This paper examines the disruption timescale distribution for
  hierarchical triples using the same methodology of Paper I, and
  addresses the following questions more specifically: (i) the
  relation between Lyapunov stability criterion and Lagrange stability
  as a function of disruption timescales of triples with different
  orbital configurations, (ii) the dependence of disruption timescales
  on the mutual inclination, $\imut$, between the inner and outer
  orbits, and (iii) the effect of the relative orbital phases among
  the three bodies that has not been explicitly studied in most
  previous literatures.

\section{Method \label{sec:method}}

We consider hierarchical triple systems comprising an inner binary
($m_1$ and $m_2$), and a tertiary ($m_3$). Since one of the main
  purposes of the present paper is to identify the mutual inclination
  dependence on the disruption timescale, we simply fix the mass ratio
  as $m_1=m_2=5m_3$, corresponding to $q_{21}\equiv m_2/m_1=1$ and
  $q_{23}\equiv m_2/m_3=5$ in Paper I.  The numerical simulation
  employs the $N$-body integrator {\tt TSUNAMI} \citep[][Trani et al.,
    in prep.]{Trani2019} that is specifically designed to accurately
  simulate few-body systems; further details of the code can be found
  in \citet{Trani2019, trani2022b} and Paper I.

We consider an initially circular inner orbit with fixing the inner
mean anomaly $M_\ii$ and the inner pericenter argument
$\omega_\ii$. We specify their explicit values for each result
below. In the barycentric reference frame whose $xy$-plane is
  chosen as the invariant plane perpendicular to its angular momentum
  vector, the inner and outer longitudes of the ascending node differ
  by $180^\circ$ (see also Fig.~1 of Paper I). Without a loss of
generality, we set them to $\Omega_\ii= 180^\circ$ and
$\Omega_\oo=0^\circ$. Therefore, the phase information of the initial
conditions is then fully specified by the values of the outer
pericenter argument and mean anomaly, $\omega_\oo$ and $M_\oo$. We fix
these values first, and vary them randomly to check the initial phase
dependence in section \ref{subsec:landscape}

We adopt the definition of the disruption time $\Td$ following Paper I
and similar to \citet{Manwadkar2020,Manwadkar2021}; at each timestep, the
integrator evaluates the binding energy for each of the three pairs of
bodies, {\it i.e.} $(m_1, m_2)$, $(m_1,m_3)$ and $(m_2,m_3)$. The pair
with the highest (negative) binding energy is considered as the inner
binary, and we call the pair of the inner binary and the remaining
tertiary as the outer pair. When the binding energy of the outer pair
becomes positive and the radial velocity of the tertiary becomes
positive (i.e. is moving away from the inner binary), we record the system as a candidate of the disrupted system tentatively. We found that the disruption normally occurs very fast after one body is ejected from a system (see also Figure 14 in Paper I).

A fraction of them, however, just represent a transient unbound state,
and become gravitationally bound again. Thus, in order to make sure
the system is truly disrupted, we continue the run until its
binary-single distance exceeds 20 times the binary semi-major axis. In
that case, we stop the run, and the triple is classified as disrupted with recording the epoch as $T_\mathrm{d}$. Otherwise, we
reset the disruption candidate condition, and keep
running the integration of the system.

Finally, we define a triple as {\it (Lagrange) stable} when $\Td
>t_\mathrm{int}$, i.e. when it does not disrupt and instead survives
until our maximum integration time $t_\mathrm{int}$. Nevertheless, we would like to emphasize again that the above definition is just for convenience, and the main purpose of the present paper is to study the disruption timescales related to the Lagrange stability.
In the present paper, 
we set
$t_\mathrm{int}= 10^{7} P_\oo$ unless otherwise specified, in contrast
to Paper I that adopts $t_\mathrm{int}= 10^{9} P_\ii$, because
  the current paper expresses the normalized disruption time
  $\Td/P_\oo$ instead of $\Td/P_\ii$. In reality, however, the result
  is identical except for the boundary of stable and unstable
  triples. We also note that under Newtonian gravity alone, the
  distribution of $\Td/P_\oo$ is scale-free (Paper I). So, if we
  consider an outer orbit of $P_\oo=10^3\,$yr, for instance, the
  stopping time of our simulations corresponds to $t_\mathrm{int}=
  10\,$Gyr, i.e., the present age of the universe. Even for
  $P_\oo=10\,$yr, $t_\mathrm{int}= 10^8\,$yr would be a practically
  reasonable timescale to consider the fate of astronomical triples.

We emphasize that the above definition of the disruption is
  expected to correspond to the Lagrange stability, instead of the
  Lyapunov stability studied by MA01. The Lyapunov {\it instability}
  may be a sufficient condition to the Lagrange {\it instability} (for
  infinite future). In practice, however, Lagrange instability is
  difficult to be identified with a limited computational time and its
  numerical accuracy. Thus we decided to examine the corresponding
  disruption timescales up to $10^7 P_\oo$, and not to repeat the
  Lyapunov-type analysis in the present paper.  We simply clarify that
  the two stabilities are different concepts, and their relation is
  not trivial in general. As we show below, Lyapunov stability, at
  least the criterion by MA01 that involves an empirical extrapolation
  of the inclination dependence, is neither a necessary nor sufficient
  condition for Lagrange stability up to $10^7 P_\oo$.

\section{Result \label{sec:result}}

\subsection{Previous models of the disruption timescale and
    Lyapunov stability criteria}

\begin{figure*}
\begin{center}
\includegraphics[clip,width=8cm]{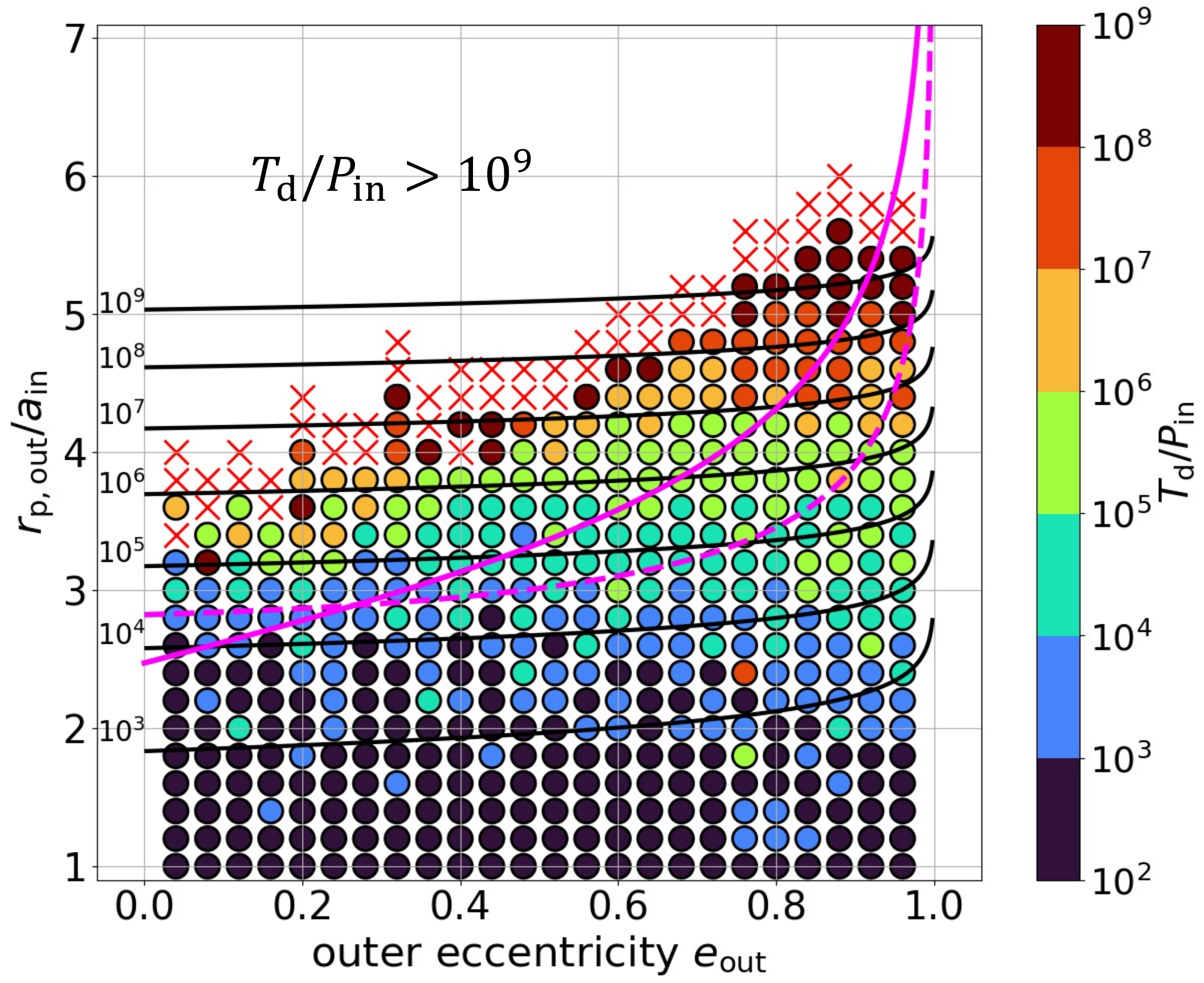}
\hspace{15pt}
\includegraphics[clip,width=8cm]{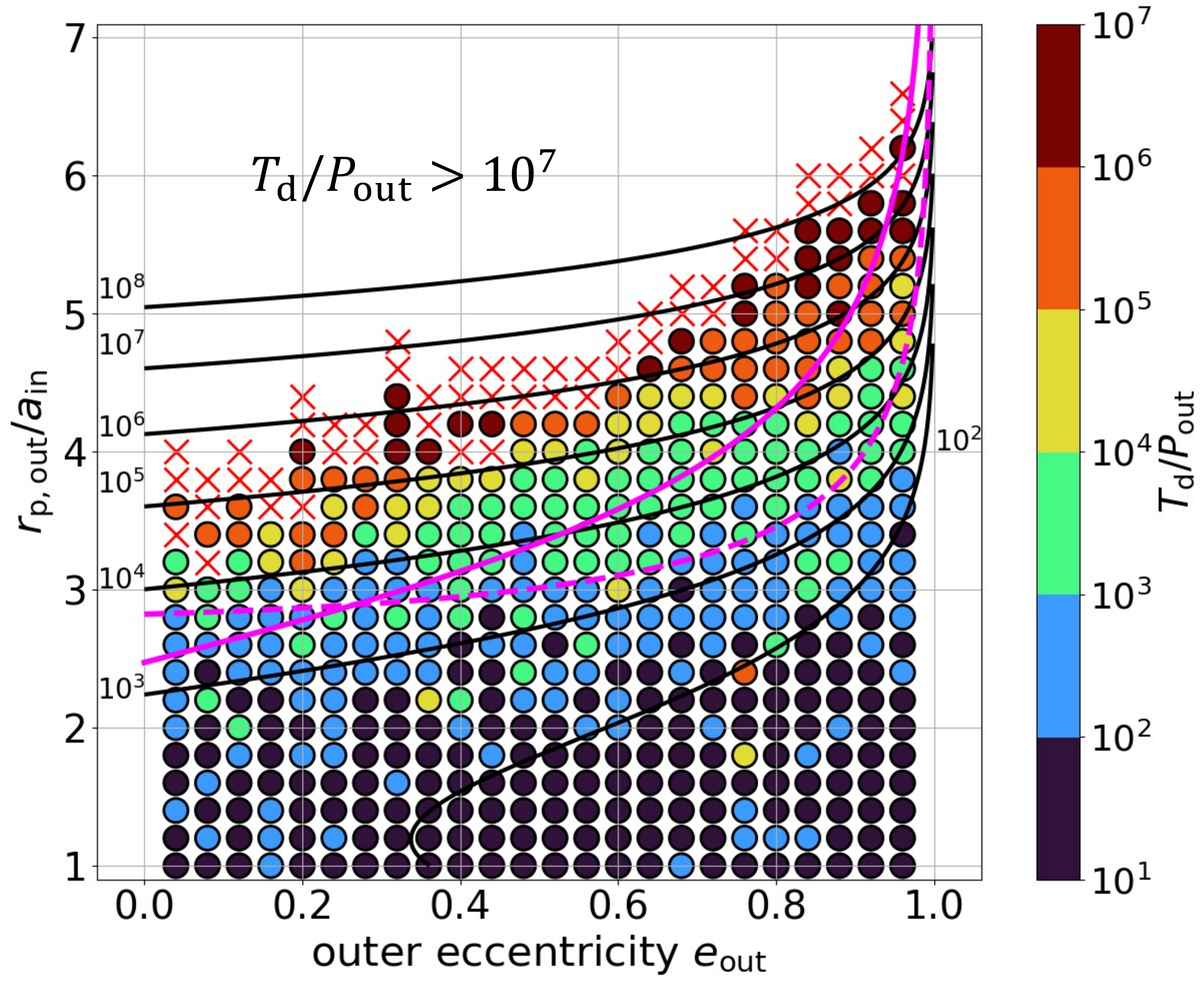}
\end{center}
\caption{Disruption timescale distribution for orthogonal triples
  ($\imut=90^\circ$) with an equal-mass inner binary ($m_1=m_2$) and
  the tertiary of $m_3=m_1/5$. We adopt
  $(\omega_\ii,M_\ii,\omega_\oo,M_\oo)=$$(180^\circ,30^\circ,0^\circ,45^\circ)$.
  The disruption timescales $T_{\rm d}$ are normalized in units of
  $P_\ii$ (left) and $P_\oo$ (right). The magenta solid and dashed
  curves represent the dynamical stability criteria,
  \citet{Mardling2001} and \citet{Vynatheya2022} (see equations
  (\ref{eq:MAcriterion}) and (\ref{eq:Vcriterion})), respectively. The
  black counters show the Random Walk model estimations of disruption
  times from \citet{Mushkin2020} (see equations (\ref{eq:rwmodel-Pin})
  and (\ref{eq:rwmodel-Pout})). Cross symbols indicate the triples
  that have not been broken until $t_\mathrm{int}$. Left and
  right plots adopt $t_\mathrm{int}=10^9P_\ii$ and
  $t_\mathrm{int}=10^7P_\oo$, respectively. We note that the left plot
  is adapted from Paper I.
\label{fig:fig1}}
\end{figure*}

In order to illustrate the basic idea behind the present
  analysis, we show in Figure \ref{fig:fig1} an example of the
  distribution of the disruption timescale $\Td$ for the simulated
  orthogonal triples on $e_\oo$ -- $\rpo/a_\ii$ plane. Here, $e_\oo$
  is the eccentricity of outer orbit, and $\rpo/a_\ii$ is the ratio of
  the pericenter distance of outer orbit and the semi-major axis of
  inner orbit.

Following Paper I, we sequentially perform the simulations at each
$e_\oo$ from lower to higher values of $\rpo/a_\ii$. Once the previous
two realizations at lower $\rpo/a_\ii$ do not disrupt before
$t_\mathrm{int}$, we stop further simulations for one sequence, and
define the boundary to save computational cost. The detail of
  this procedure is described in Paper I.  The left and right panels
in Figure \ref{fig:fig1} plot $\Td/P_\ii$ and $\Td/P_\oo$,
respectively. While the information context of the two panels is
essentially the same, their comparison provides a useful insight into
the purpose of the present paper.

For reference, we overlay the disruption timescale contours predicted
from the random walk (RM) model by \citet[][]{Mushkin2020} in black
solid lines:
\begin{eqnarray}
\label{eq:rwmodel-Pin}
\frac{\Td}{P_\ii}
= 2\left(\frac{m_{123}m_{12}}{m_1 m_2}\right)^2\sqrt{1-e_\oo}
\left(\frac{\rpo}{\ain}\right)^{-2}
\exp\left[\frac{4\sqrt{2}}{3}\sqrt{\frac{m_{12}}{m_{123}}}
\left(\frac{\rpo}{\ain}\right)^{3/2}\right],
\end{eqnarray}
and
\begin{eqnarray}
\label{eq:rwmodel-Pout}
\frac{\Td}{P_\oo}
= 2\left(\frac{m_{12}^2}{m_1 m_2}\right)^2
\left(\frac{m_{123}}{m_{12}}\right)^{5/2}(1-e_\oo)^2
\left(\frac{\rpo}{\ain}\right)^{-7/2}
\exp\left[\frac{4\sqrt{2}}{3}\sqrt{\frac{m_{12}}{m_{123}}}
\left(\frac{\rpo}{\ain}\right)^{3/2}\right],
\end{eqnarray}
where $m_{123}\equiv m_1+m_2 +m_3$ is the total mass of the system,
and $m_{12}\equiv m_1+m_2$ is the total mass of inner binary.
Equations (\ref{eq:rwmodel-Pin}) and (\ref{eq:rwmodel-Pout})
  serve as useful analytical estimates for the disruption timescale
  of Lagrange-unstable systems.

In Figure \ref{fig:fig1}, we also include the dynamical stability
criteria by MA01 and \citet{Vynatheya2022} as magenta solid and dashed
lines, respectively.  \citet{Vynatheya2022} defined the boundary from
their simulations at $100P_\oo$ by selecting the triples whose inner
and outer semi-major axes do not change by more than 10 percent of the
initial values. This is not exactly the same condition adopted by MA01, but
is expected to be similar from the viewpoint of Lyapunov stability.

According to them, those triples with $\rpo/\ain$ exceeding the
  critical thresholds below are Lyapunov stable, respectively:
\begin{eqnarray}
\label{eq:MAcriterion}
 \left(\frac{\rpo}{\ain}\right)_\mathrm{MA}
	\equiv 2.8\left(1-0.3\frac{i_\mathrm{mut}}{180^\circ}\right)
	\left[\left(1+q_\oo\right)
	\frac{(1+e_\oo)}{\sqrt{1-e_\oo}}\right]^{2/5}
\end{eqnarray}
and
\begin{eqnarray}
\label{eq:Vcriterion}
\left(\frac{\rpo}{\ain}\right)_\mathrm{V}
&\equiv& 2.4 (1+\tilde{e}_\ii)\left[
\frac{(1+q_\oo)}{\sqrt{1-e_\oo}(1+\tilde{e}_\ii)}\right]^{2/5} \nonumber \\
&\times& \left[\left(\frac{1-0.2\tilde{e}_\ii+e_\oo}{8}\right)(\cos{i_\mathrm{mut}}-1)+1\right],
\end{eqnarray}
where $q_\oo\equiv m_3/m_{12}$, and $\tilde{e}_\ii$ is defined as 
\begin{eqnarray}
\label{eq:etilde}
\tilde{e}_\ii \equiv
\max\left(e_\ii, 0.5\left(1-\frac{5}{3}\cos^2{i_\mathrm{mut}}\right)\right).
\end{eqnarray}

The left panel of Figure \ref{fig:fig1} is adapted from the simulation
results with $t_\mathrm{int}=10^9~P_\ii$ in Paper I. Due to the
chaotic nature of three-body dynamics, the distribution of $\Td$
exhibits relatively large scatters; see Figures 6, 7, 16, 17 and
  18 in Paper I. Nevertheless, the values of $\Td/P_\ii$ are
basically determined by the value of $\rpo/\ain$.  This is consistent
with the RW model prediction, equation (\ref{eq:rwmodel-Pin}), which
suggests that $\Td/P_\ii$ depends on $e_\oo$ very weakly ($\propto
\sqrt{1-e_\oo}$).  The right panel shows $\Td/P_\oo$ from our new
simulations with $t_\mathrm{int}=10^7~P_\oo$.

Since equations (\ref{eq:MAcriterion}) and (\ref{eq:Vcriterion})
  represent the thresholds for Lyapunov stability, there is no reason
  why their criteria are directly related to the disruption timescale
  characterizing Lagrange stability. Figure~\ref{fig:fig1} indicates
  that their Lyapunov stability boundaries indeed correspond to a
  broad range of disruption timescales, $10^2<\Td/P_\oo<10^5$ in
  our simulation runs. In other words, their criteria should not be
  interpreted as a {\it black-and-white} boundary of the disruption of
  triples. This seems to be a common misinterpretation in works
  studying triple stability (R. Mardling, private communication). This
  is why we have repeatedly stressed the conceptual difference between
  Lagrange and Lyapunov stabilities.

The comparison of the two panels in Figure \ref{fig:fig1}
  motivates us to study more systematically the dependence of
  $\Td/P_\oo$ on the mutual inclination $\imut$ of the inner and outer
  orbits. We note again that Paper~I adopted the normalized disruption
  timescales of $\Td/P_\ii$, while the present paper uses
  $\Td/P_\oo$. This is not essential and just a matter of definition,
  but the different normalization may be useful in providing a
  complementary view to the problem relative to Paper I. Furthermore,
  if the outer binary is highly eccentric, most of the energy transfer
  between the two orbits, which is responsible for triggering the
  instability, should occur once every $P_\oo$ when the
  outer body is at pericenter. Therefore, the scaling of $\Td/P_\oo$
  may be more relevant from the physical perspective.

\subsection{Mutual inclination dependence of the disruption timescales}

\begin{figure*}
\begin{center}
\includegraphics[clip,width=14.5cm]{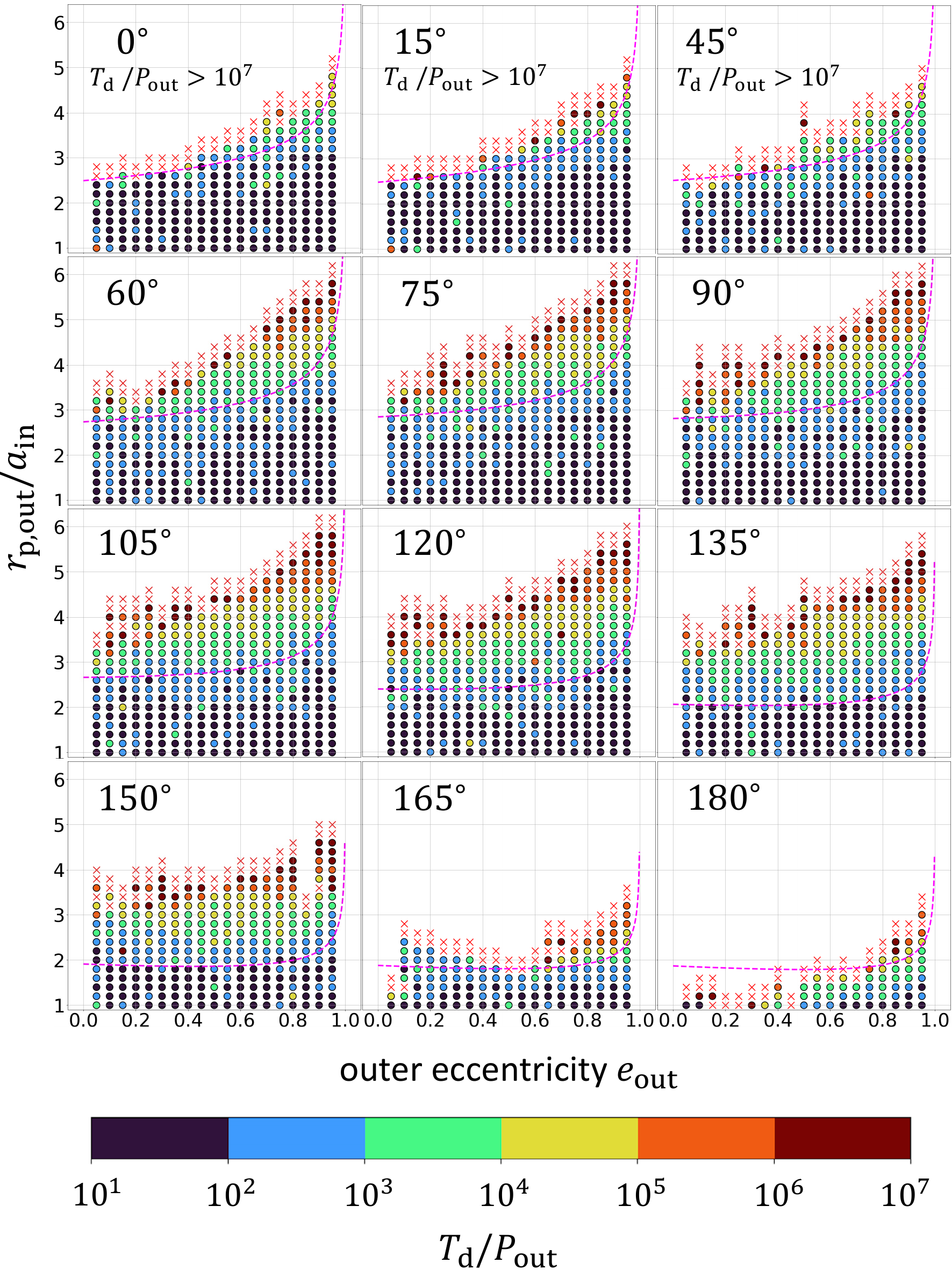}
\end{center}
\caption{The normalized disruption timescales $T_\mathrm{d}/P_\oo$ on
  $e_\oo - \rpo/a_\ii$ plane for different values of $\imut$.  The
  disruption timescales evaluated from simulations are indicated
  according to the side color scales. Magenta curves represent the
  dynamical stability criterion from \citet{Vynatheya2022} (see
  equation (\ref{eq:Vcriterion})). Cross symbols indicate the triples
  that have not been broken until $t_\mathrm{int}=10^7
  P_\oo$. While the result for $\imut=30^\circ$ is not shown here, it
  is very similar to a simple interpolation of those for
  $\imut=15^\circ$ and $45^\circ$.
\label{fig:fig2}}
\end{figure*}

Figure~\ref{fig:fig2} plots $T_\mathrm{d}/P_\oo$ from our simulations
on the $e_\oo - \rpo/a_\ii$ plane for different values of
$\imut$. This generalizes Paper I that focused on three representative
mutual inclinations ($i_\mathrm{mut}=0^\circ$, $90^\circ$ and
$180^\circ$) alone, and quantitatively estimates the disruption
  timescales associated with Lagrange stability by systematically
  varying the initial inclinations $\imut$ in an interval of
  $15^\circ$.  We adopt the same initial parameter sets in Paper I and
  Figure \ref{fig:fig1}, and fix the initial phases as
  $(\omega_\ii,M_\ii, \omega_\oo, M_\oo)
  =(180^\circ,30^\circ,0^\circ,45^\circ)$. The estimated timescales
  are supposed to vary by one or two orders of magnitude when the
  initial phases are randomly assigned, in addition to the similar
  amount of scatters due to the intrinsic chaotic nature of the triple
  dynamics (Paper I). Thus, the distribution of $T_\mathrm{d}/P_\oo$
  shown in Figure~\ref{fig:fig2} should be understood to have a
  scatter of one or two orders of magnitude in general.  The variation
  due to the different initial phases will be examined in
  Figure \ref{fig:fig3} below.

For reference, we plot the stability boundary proposed by
  \citet{Vynatheya2022} alone in Figure~\ref{fig:fig2}; as
  Figure~\ref{fig:fig1} indicates, the MA01 boundary is roughly
  identical. The top three panels indicate that triples with
  $\rpo/\ain < (\rpo/\ain)_{\rm V}$ are disrupted approximately within
  $\Td<10^3 P_\oo$, while those with $\rpo/\ain > (\rpo/\ain)_{\rm V}$
  mostly survive more than $10^7 P_\oo$.  In that sense, Lyapunov
  stable triples with $\imut<45^\circ$ (nearly coplanar prograde) are
  Lagrange stable as well.

However, it is not the case in general.  In particular, strong
  ZKL oscillations for $\imut>60^\circ$ tend to destabilize the
  triple, and the fate of those triples can be revealed only by
  integrating them much longer than the corresponding quadrupole ZKL
  timescale \citep[e.g.][]{Antognini2015}.
\begin{eqnarray}
\label{eq:tau-ZKL}
\frac{T_{\rm ZKL}}{P_\oo} = \frac{m_{123}P_\oo}{m_3 P_\ii} (1-e_\oo^2)^{3/2}
=\frac{m_{12}}{m_3}\sqrt{\frac{m_{123}}{m_{12}}}
\left(\frac{\rpo}{\ain}\right)^{3/2}(1+e_\oo)^{3/2}.
\end{eqnarray}
Indeed, the disruption timescales for triples with
$60^\circ<\imut<150^\circ$ monotonically increase as $\rpo/\ain$
without any discontinuous change around $\rpo/\ain = (\rpo/\ain)_{\rm
  V}$. This fact implies that the Lagrange instability is mainly triggered
via the repeated ZKL oscillations over long timescales.

Furthermore, the significant suppression of the energy transfer
between the inner and outer orbits for $\imut \approx 180^\circ$
strongly stabilize those triples in coplanar retrograde orbits (Paper
I) from the viewpoint of disruption timescales. Thus an accurate prediction of their fate requires a much
longer-time integration as well.

Figure~\ref{fig:fig2} clearly illustrates the impact of $\imut$ on the
disruption timescale of triples. In particular, it is worth noting
that the Lagrange stability boundary appears to be highly asymmetric
with respect to $\imut=90^\circ$. This is not expected from the
quadrupole ZKL interaction mechanism alone, because it should be
symmetric around $90^\circ$. The broken symmetry might arise due to
semi-secular effects that excite the eccentricity of the inner binary
more efficiently than the simple quadrupole interaction mechanism
\citep{Grishin2018,Mangipudi2022}, and thus increase the energy
diffusion between inner and outer orbits (Paper I).

\subsection{Dependence of disruption timescales on
    \texorpdfstring{$\rpo/\ain$, $e_\oo$ and $\imut$}{rp/ain, eout and
      imut} with randomized initial phases \label{subsec:landscape}}

Due to the chaotic nature of the triple dynamics, the result of Figure
\ref{fig:fig2} may change significantly by the difference of the
initial phases (see Paper I). Therefore, we examine the sensitivity of
the disruption timescales to the initial phases by computing sections
of Figure \ref{fig:fig2} along the constant $e_\oo$ direction. To be
specific, we fix $M_\ii=0^\circ$ and $\omega_\ii = 0^\circ$, and
change $\rpo/a_\ii$ by a step of $0.05$, which is four times denser
sampling than those adopted in Figure \ref{fig:fig2}.  Furthermore, we
compute 10 different realizations for the same set of parameters,
$(e_\oo, \rpo/\ain, \imut)$ using randomly selected values of
$\omega_\oo$ and $M_\oo$. We emphasize that the tiny difference
  of other orbital parameters induces the scatters of $\Td/P_\oo$ by a
  similar amount (see Paper I).

\begin{figure*}
\begin{center}
\includegraphics[clip,width=9cm]{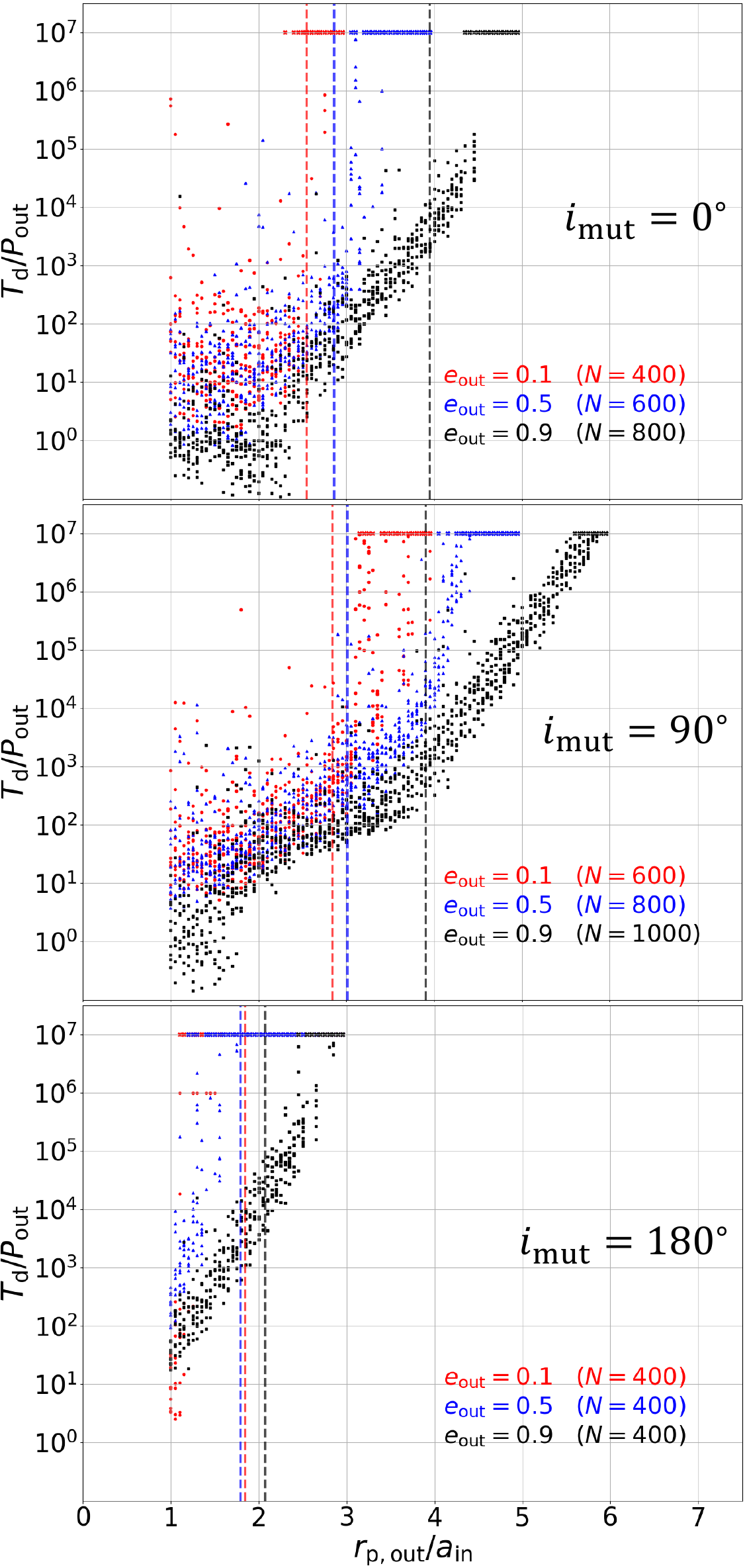}
\end{center}
\caption{Behavior of the stable and unstable transition for triples
  with $\imut=0^\circ, 90^\circ$, and $180^\circ$. In each panel, the
  different colored symbols plot $T_{\rm d}/P_\oo$ against
  $\rpo/a_\ii$ for $e_\oo=0.1$ (red), $0.5$ (blue), and $0.9$ (black).
  For each set of $(e_\oo, \rpo/\ain, \imut)$, we run 10 realizations
  with different initial phases $(\omega_\oo,M_\oo)$ randomly sampled
  from uniform distribution. The value of $N$ in the panels denotes
  the total number of the corresponding runs (data points).
\label{fig:fig3}}
\end{figure*}

Figure \ref{fig:fig3} shows the resulting $T_\mathrm{d}/P_\oo$
distributions for coplanar prograde (top), orthogonal (middle), and
coplanar retrograde (bottom) triples. Just for reference, we plot the
expected stability boundary, equation (\ref{eq:Vcriterion}), in dashed
lines. The top panel shows that $T_\mathrm{d}/P_\oo$ is a very
  sensitive function of $\rpo/\ain$ for low-inclined systems, even
  taking into account the scatters due to initial phase differences;
$T_\mathrm{d}/P_\oo$ changes significantly around the transition
region corresponding to the value of $10^4$--$10^5$. The top panel
have very few data points between $10^5$--$10^7$, while a majority of
triples are simply located at the upper limit of integration time
($T_{\rm d}/P_\oo= 10^7$).  This result agrees qualitatively with that
of \citet{Vynatheya2022} except for highly eccentric cases. For a
highly eccentric case of $e_\oo = 0.9$, the transition becomes less
abrupt; $T_\mathrm{d}/P_\oo$ increases gradually with $\rpo/a_\ii$ up
to $4.5$, and then shows significant jump.

The bottom panel, for coplanar retrograde triples, shows a similar
behavior, although the systems are significantly more stable than
prograde cases, and the transition looks less abrupt.

In contrast, for orthogonal triples (middle panel);
$T_\mathrm{d}/P_\oo$ increases gradually with $\rpo/a_\ii$ until
$\rpo/a_\ii \approx 3$ and $4$ for $e_\oo=0.1$ and $0.5$,
respectively, and then exhibits significant increase only when
$\rpo/a_\ii$ exceeds those values. Even after exceeding those values,
the transition is still not so abrupt for orthogonal triples, unlike
the case for coplanar prograde triples.  For $e_\oo=0.9$,
$T_\mathrm{d}/P_\oo$ continues to increase gradually with $\rpo/a_\ii$
up to $\rpo/a_\ii=6$ where $T_\mathrm{d}/P_\oo$ reaches our upper
limit of the integration time ($t_\mathrm{int}=10^7P_\oo$). Those
trends are statistically robust against the different choices of the
initial phases as illustrated by the scatters of the data points.

\section{Summary and conclusion \label{sec:conclusion}}

We have computed the distribution of disruption timescales for
hierarchical triples with an equal-mass and initially circular inner
binary, using a series of N-body simulations.  We extend
\citet{HTS2022} (Paper I), and systematically examine the dependence
on the mutual inclination between the inner and outer orbits in
particular. By integrating the triple systems much longer than the
previous studies, we are able to reveal the fate of those triples more
reliably than before. Our main findings are summarized as follows.

\begin{itemize}
\item[(1)] In studying the dynamical stability of hierarchical
  triple systems, it is important to distinguish between Lyapunov and
  Lagrange stabilities. A widely used stability criterion by
  \citet{Mardling1999, Mardling2001} and \citet{Mardling2008} is
  derived from the local divergence of the orbits, and thus
  corresponds to the former.  Paper I and the present paper compute
  the disruption timescale of triple that is related to the Lagrange
  stability. Those concepts are complementary but very different,
  which should be kept in mind when applying those results for
  specific purposes.
\item[(2)] Dynamical disruption timescales of triples are very
  sensitive to the mutual inclination $\imut$ of the inner and outer
  orbits.  They change drastically around the Lyapunov stability
  boundary if $\imut<45^\circ$. Thus, the Lyapunov {\it stability}
  condition also guarantees the long-term Lagrange {\it stability} for
  moderately inclined triples.  In contrast, for triples of
  significantly inclined inner and outer orbits ($60^\circ < \imut <
  150^\circ$), the disruption timescales vary smoothly against
  $\rpo/a_\ii$ due to the ZKL oscillations.  The difference of the
  initial phases adds one or two orders of magnitudes scatters to the
  disruption timescales, but does not change the behavior in a
  statistical sense.
\item[(3)] The Lagrange stability of triples does not change
  monotonically with the mutual inclination. For triples with moderate
  inclinations ($\imut <45^\circ$), the stability is not so sensitive
  to $\imut$.  By increasing $\imut$ up to $\sim 100^\circ$, those
  triples become less stable due to the ZKL oscillations. Retrograde
  triples with $\imut>120^\circ$ become stabilized again as $\imut$
  increases, and the coplanar retrograde triple ($\imut=180^\circ$) is
  the most stable configuration.  \citet{Grishin2017} found the
  similar behavior about the mutual inclination dependence in the Hill
  stability, and pointed out that the ZKL oscillations affect the
  stability.  Interestingly, the above behavior is not symmetric
  around $\imut=90^\circ$, in contrast to a naive expectation from the
  quadrupole ZKL oscillations, indicating the importance of
  quasi-secular effects \citep[e.g.][]{Grishin2018,Mangipudi2022}
  beyond the simple ZKL quadrupole interaction.
\end{itemize}

The present paper has adopted a range of parameters in which the
Newtonian gravity is dominant and the general relativity (GR)
corrections are negligible. However, it is known that the GR
precessions reduce the ZKL effects for specific parameter ranges
\citep[e.g.][]{Liu2015}. In this case, the Lagrange stability for
inclined cases may be highly affected by GR corrections. In addition,
gravitational wave emissions would reduce the semi-major axis of the
inner binary, and thus affect the stability of triples with inner
compact objects.

In this paper, we also fix the mass ratios, $q_{21}\equiv
  m_2/m_1=1$ and $q_{23}\equiv m_2/m_3 = 5$. Different choices of
  those parameters are important in applying to a variety of
  interesting astronomical objects. For instance, small $q_{21}$ and
  $q_{23}$ are relevant for planetary systems. In addition, octupole
  ZKL oscillations play an important role if $q_{21}$ deviates
  significantly from unity, {\it i.e.} for a very unequal mass inner
  binary. The strong enhancement of the eccentricity of the inner
  binary by the octupole-level ZKL interactions would affect the
  stability.

Many scenarios are proposed recently to explain a variety of
astronomical phenomena via the ZKL oscillations, including the
formation of extrasolar hot jupiters \citep[e.g.][]{Wu2003,Naoz2012},
and the acceleration of compact mergers
\citep[e.g.][]{Liu2018,Trani2021}. We show that the concept of the
disruption timescales is particularly more important for triples with
strong ZKL oscillations. Therefore, quantitative implications of these
scenarios have to be discussed with taking account the disruption
timescales.

Generalizing the current stability analysis of triples by taking
account of the problems above-mentioned is an important area of
research, which we plan to address in due course.

\section*{Acknowledgments}

We thank the referee of the paper, Rosemary Mardling, for various
critical and insightful comments. In particular, she pointed out the
important difference between the chaoticity and Lagrange instability
of triple systems, which had not been explicitly described in the
earlier manuscript of the paper. We would like to add that the referee
and we do not yet fully agree upon the relation between the
dynamically chaotic triples and the Lagrange unstable
triples. Nevertheless, the referee agreed that it should be a subtle
point which is beyond the scope of the present paper and deserves
future investigations. In any case, we do appreciate her invaluable
comments and suggestions, which significantly improved the conceptual
clarity of the paper.  We are also grateful to Evgeni Grishin for
useful discussion, and Pavan Vynatheya for correspondences concerning
the detail of the stability criterion in his work. T.H. acknowledges
the fellowship by Japan Society for the Promotion of Science
(JSPS). This work is supported partly by the JSPS KAKENHI grant
Nos. JP18H01247 and JP19H01947 (Y.S.), JP21J11378 (T.H.), and
JP21K13914 (A.A.T).  The numerical simulations were carried out on the
local computer cluster \texttt{awamori} purchased from those grants.

\end{document}